# Relations between semantic security and indistinguishability against cpa, non-adaptive cca and adaptive cca in comparison based framework

Ali Bagherzandi          Javad Mohajeri         Mahmoud Salmasizadeh
bagherzandi@ce.sharif.edu    Mohajer@sharif.edu      salmasi@sharif.edu

**Abstract**. In this paper we try to unify the frameworks of definitions of semantic security, indistinguishability and non-malleability by defining semantic security in comparison based framework. This facilitates the study of relations among these goals against different attack models and makes the proof of the equivalence of semantic security and indistinguishability easier and more understandable. Besides, our proof of the equivalence of semantic security and indistinguishability does not need any intermediate goals such as non devidability to change the definition framework.

## 1 Introduction

The security of public key cryptosystems can be evaluated as achieving certain cryptographic goals such as semantic security, indistinguishability, non-malleability, plaintext awareness and non-devidability. In this paper we focus on semantic security and indistinguishability which has been defined in [GM84] for the first time. The latter is also known as polynomial security or Goldwasser-Micali security.

Roughly speaking, indistinguishability formalizes an adversary's inability to distinguish between two plaintexts given the encryption of one of them. It is rather an artificial goal but suggests an applicable method for evaluating security in provable security context. On the other hand, an encryption scheme is said to be semantically secure if no polynomially bounded adversary can be found to extract any partial information about the plaintext of a given ciphertext. Thus semantic security is a direct intuition of privacy and comparing whit Shannon's perfect security [S49] it can be considered as the computational version of perfect security. Unlike indistinguishability, semantic security does not suggest any method for security evaluation.

The term "information" in the definition of semantic security can be modeled by functions from message space to $S^*$ in which $S$ is the alphabet of computation model. Proving that no such function exists for a cryptosystem implicitly proves the indistinguishability goal for that cryptosystem. Such a close relationship between indistinguishability and semantic security was firstly demonstrated in [GM84] as their equivalence. In the original definition of semantic security in [GM84] there is no restriction on the computability of the functions modeling information about plaintext. But as it has bean said in [MRS88] what good would it do any adversary to "guess" a function if he can not even verify that his guess is correct. In later formulations of semantic security the functions modeling "information" restricted to be polynomially verifiable [WSI03].

Another important turning point was introduced in [BDPR98]. Bellare et al. suggested that cryptographic goals to be studied in connection with attack models and not in isolation. Using this method, relations among indistinguishability and semantic security is discussed in [BDPR98] against chosen plaintext attack, non-adaptive chosen ciphertext attack and adaptive chosen ciphertext attack.

Semantic security can be formalized under two different frameworks namely simulator based and comparison based. The simulator based definitions requests that for any adversary given a ciphertext there exists a poly-time algorithm called a simulator which succeeds in the attack (i.e. extracting non negligible information) without the ciphertext essentially as well as the adversary. The comparison based definition requests that any adversary in possession of the ciphertext obtains no advantage over one which performs only random guess. Since random guess can be regarded as a special case of simulation the comparison based notion may seem stronger than the simulator-based one. On the other hand in the simulator based definition there is no restriction on the computability of partial information which an adversary whishes to extract while in the comparison based the partial information has to be efficiently generated and evaluated by a poly-time algorithm. This may show that the former is stronger than the latter.

In [BDPR98] non-malleability is defined in comparison based framework while in previous definitions [DDN95, DDN98] it has bean defined using a simulator. Besides in [BDPR98] it has been shown that these two definitions are equivalent. Semantic security as defined in [WSI03] is a simulator based one. Using this definition results in some contradictions as mentioned above and makes proving the equivalence between semantic security and indistinguishability difficult. Besides, the proof of the equivalence between semantic security and indistinguishability in this framework requires some other artificial security goals such as non-devidability to be defined in order to unify the definition frameworks.

In this paper using the idea in [BDPR98], we define semantic security in comparison based framework and study the relations between semantic security and indistinguishability against chosen plaintext attack ($cpa$), non-adaptive chosen ciphertext attack ($cca1$) and adaptive chosen ciphertext attack ($cca2$). Finally we suggest a simple and more understandable proof of equivalence of semantic security and indistinguishability.

Figure (1) shows the summary of works done in this category.

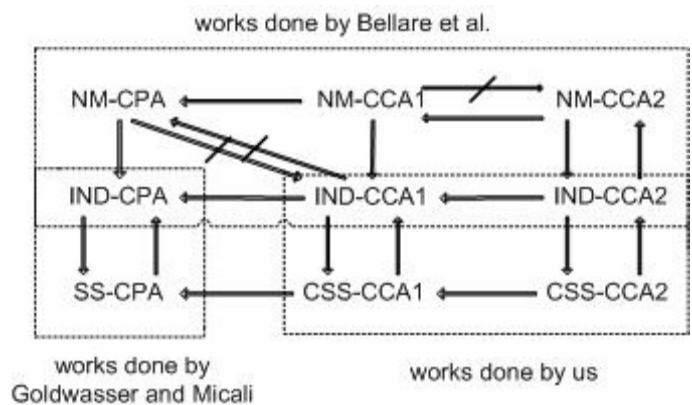

figure (1) Relations among security notions

In figure (1) the arrows indicate implication and the hatched arrows indicate the non implications. Thus the existence of a path from a pair of goal – attack $G_1 - A_1$ to $G_2 - A_2$ shows that if the goal $G_1$ in a cryptosystem is achieved in the sense of attack $A_1$ then goal $G_2$ is also achieved against attack $A_2$. For example if a cryptosystem is proved to be non-malleable against adaptive chosen ciphertext attack then it is semantically secure against chosen plaintext attack.

In section 2 we introduce some preliminary definitions. Then in section 3 we introduce our definition of semantic security as well as a slightly modified definition of indistinguishability based on comparison and then in section 4 we study our proof of equivalence of semantic security and indistinguishability in the new framework.

## 2 Preliminary definitions

**Polynomially verifiable function**

The function $f : M \to S^*$ is said to be polynomially verifiable if there exists a probabilistic poly-time algorithm $A$ such that:
$\forall x \in M : A(x, f(x)) = 1$ and $y \neq f(x) \Rightarrow A(x, y) = 0$

In this definition $M$ is the message space and $S^*$ is the whole information about plaintext.

**Negligible function**

The function $e : N \to R$ is negligible if
(1) $\forall n \in N : e(n) \geq 0$
(2) $\forall c \geq 0 \exists k_c \ni \forall k \geq k_c e(k) < k^{-c}$

**Public key encryption scheme** A public key encryption scheme $P = (K, E, D)$ is a triple of algorithms such that:
(1) The key generation algorithm $K$ is a probabilistic poly-time algorithm that takes a security parameter $k \in N$ as the input and outputs a pair $(pk, sk)$ of matching public and secret keys,
(2) The encryption algorithm $E$ is a probabilistic poly-time algorithm that takes a public key $pk$ and a message $x \in \{0,1\}^*$ as the input and outputs a ciphertext $y$,
(3) The decryption algorithm $D$ is a deterministic poly-time algorithm that takes a secret key $sk$ and a ciphertext $y$ as the input and outputs either a message $x \in \{0,1\}^*$ or a special symbol $o$, if no legal decryption can be found for $y$.

**Adversary Model**

An adversary is modeled as a pair of probabilistic poly-time algorithm $A = (A_1, A_2)$. The exact purpose of each algorithm depends on the particular adversarial goal, but in general, in the first stage i.e. using $A_1$, the adversary given the public key seeks and outputs some test instance and in second stage the adversary is issued a challenge ciphertext $y$ generated as a probabilistic function of the test instance in a manner depending on the goal. In addition, $A_1$ can output some state information that will be passed to $A_2$. The adversary $A = (A_1, A_2)$ is said to be successful if it passes the challenge.

In chosen plaintext attack ($CPA$) the adversary can encrypt any arbitrary plaintext

In non-adaptive chosen ciphertext attack ($CCA1$) we give $A_1$ (the public key) and access to a decryption oracle but we do not allow $A_2$ access to a decryption oracle. Thus the decryption oracle can be used to generate test instance but is taken away before the challenge appears.

In adaptive chosen ciphertext attack ($CCA2$) we continue to give $A_1$ the public key and the access to decryption oracle but we also give $A_2$ the access to the same decryption oracle, with the only restriction that the challenge ciphertext cannot be queried.

# 3 Goal definitions

**Comparison based semantic security**

Let $A = (A_1, A_2)$ be an adversary attacking the encryption scheme $P = (K, E, D)$. The adversary in the first phase of attack i.e. using algorithm $A_1$ takes as the input the public key $pk$ and outputs the pair $(M, s)$ in which the first component is a message space samplable in poly-time and the second component is any information that should be delivered from $A_1$ to $A_2$. Then a random message $x \in M$ is selected and encrypted by $E_{pk}$ to produces the ciphertext $y$. In the second phase of attack, the algorithm $A_2$ takes as the input the massage space, the state information and the challenge ciphertext i.e. $(M, s, y)$ and outputs the pair $(v, f)$ and a random $x \in M$ is selected by the algorithm *sample* using the information delivered to $A_2$.

The algorithm *sample* is said to be successful if the random $x$ that it selects satisfies the equation $v = f(x)$.

If the difference of the success of the adversary and the algorithm *sample* as a function of $k$, the security parameter, is a negligible function, then $P = (K, E, D)$ is said to be secure in the sense of $CSS\_ATK$.

In formal setting let $P = (K, E, D)$ be an encryption scheme and $A = (A_1, A_2)$ be a polynomially bounded adversary. For $atk \in \{cpa, cca1, cca2\}$ and $k \in N$, we define the experiments $Exp_{P,A,Sample}^{css-atk-1}(k)$ and $Exp_{P,A,Sample}^{css-atk-0}(k)$ as below.

$Exp_{P,A,Sample}^{css-atk-1}(k)$

    $(pk, sk) \leftarrow K(k); (M, s) \leftarrow A_1^{O_1(.)}(pk); x \leftarrow M; y \leftarrow E_{pk}(x);$

    $(v, f) \leftarrow A_2^{O_2(.)}(M, s, y);$

    if $v = f(x)$ then $d \leftarrow 1$

    else $d \leftarrow 0;$

return $d$

$Exp_{P,A,Sample}^{css-atk-0}(k)$

    $(pk, sk) \leftarrow K(k); (M, s) \leftarrow A_1^{O_1(.)}(pk); x \leftarrow Sample(M, s);$

    $(v, f) \leftarrow A_2^{O_2(.)}(M, s);$

    if $v = f(x)$ then $d \leftarrow 1$

    else $d \leftarrow 0;$

return $d$

In which for any $x, x' \in M$ we have $|x| = |x'|$; and the adversary $A$ has the oracle access to a decryption oracle as below:

if $atk = cpa$ then $O_1(\cdot) = e$ and $O_2(\cdot) = e$,

if $atk = cca1$ then $O_1(\cdot) = D_{sk}(\cdot)$ and $O_2(\cdot) = e$,

if $atk = cca2$ then $O_1(\cdot) = D_{sk}(\cdot)$ and $O_2(\cdot) = D_{sk}(\cdot)$

but in the case of $atk = cca2$, $A_2$ is not allowed to request the decryption of the challenged ciphertext $y$.

The advantage of the adversary, $Adv_{P,A,A'}^{css-atk}(k)$, is defined to be

$$Adv_{P,A,Sample}^{css-atk}(k) = Pr[Exp_{P,A,Sample}^{css-atk-1}(k) = 1] - Pr[Exp_{P,A,Sample}^{css-atk-0}(k) = 1]$$

The public key encryption scheme $P = (K, E, D)$ is secure in the sense of $CSS\_ATK$ if

$\forall A \; \exists S : Adv_{P,A,Sample}^{css-atk}(k)$ is negligible.

We can unify the two experiments above in the following experiment:

For $atk \in \{cpa, cca1, cca2\}$ and $b \in \{0,1\}$ and $k \in N$ we define

$Exp_{P,A,Sample}^{css-atk-b}(k)$

$\quad (pk, sk) \leftarrow K(k); (M, s) \leftarrow A_1^{O_1(\cdot)}(pk); x_1 \leftarrow M; y \leftarrow E_{pk}(x_1); x_0 \leftarrow Sample(M, s);$

$\quad (v, f) \leftarrow A_2^{O_2(\cdot)}(M, s, y);$

$\quad if \; v = f(x_b) \; then \; d \leftarrow 1;$

$\quad else \; d \leftarrow 0;$

return $d$

**Indistinguishability**

Let $P = (K, E, D)$ be an encryption scheme and $A = (A_1, A_2)$ be a polynomially bounded adversary. For $atk \in \{cpa, cca1, cca2\}$ and $k \in N$ and $b \in \{0,1\}$, we define $Exp_{P,A}^{ind-atk-b}(k)$ as below:

$Exp_{P,A}^{ind-atk-b}(k)$

$\quad (pk, sk) \leftarrow K(k); (x_0, x_1, s) \leftarrow A_1^{O_1(\cdot)}(pk); y \leftarrow E_{pk}(x_b);$

$\quad d \leftarrow A_2^{O_2(\cdot)}(x_0, x_1, s, y);$

return $d$

In which for any $x, x' \in M$ we have $|x| = |x'|$; and the adversary $A$ has the oracle access to a decryption oracle as below:

if $atk = cpa$ then $O_1(\cdot) = e$ and $O_2(\cdot) = e$,

if $atk = cca1$ then $O_1(\cdot) = D_{sk}(\cdot)$ and $O_2(\cdot) = e$,

if $atk = cca2$ then $O_1(\cdot) = D_{sk}(\cdot)$ and $O_2(\cdot) = D_{sk}(\cdot)$

but in the case of $atk = cca2$, $A_2$ is not allowed to request the decryption of the challenged ciphertext $y$.

The advantage of the adversary which is a criterion of its correct guess is defined as the difference of the probability of outputting a correct $1$ and a wrong $1$.

In a formal setting the advantage of the adversary is formulated as below:

$$Adv_{P,A}^{ind-atk}(k) = Pr[Exp_{P,A}^{ind-atk-1}(k) = 1] - Pr[Exp_{P,A}^{ind-atk-0}(k) = 1]$$

The public key encryption scheme $P = (K, E, D)$ is said to be secure in the sense of *IND_ATK* if

$\forall A \ Adv_{PE,A}^{ind-atk}(k)$ *is negligible*.

## 4 Equivalence between semantic security and indistinguishability

**Theorem 1:**
The public key encryption scheme $P = (K, E, D)$ is secure in the sense of *CSS_ATK* if and only if it is secure in the sense of *IND_ATK*, for any attack $ATK \in \{CPA, CCA1, CCA2\}$.

**Proof:**
First we prove the "if" part of the theorem, namely $CSS\_ATK \Rightarrow IND\_ATK$:

Suppose $P = (K, E, D)$ is a public key encryption scheme that is secure in the sense of *CSS_ATK* but it is not secure in the sense of *IND_ATK*. Thus, according to the definition of indistinguishability (section 3) there exists a poly-time adversary $A = (A_1, A_2)$ being able to distinguish between ciphertexts of some plaintexts from message space. Using $A = (A_1, A_2)$ as a subroutine, we construct the poly-time adversary $B = (B_1, B_2)$ that can extract some non-negligible information about plaintext from its corresponding ciphertext.

Constructing $B$ is straight forward; every poly-time algorithm that can distinguish between ciphertexts $x_0, x_1 \in M$, will predict the value of the following function:

$f : \{x_0, x_1\} \rightarrow \{0,1\}$

$f(x) = \begin{cases} 0 & \text{if } x = x_0 \\ 1 & \text{if } x = x_1 \end{cases}$

Thus the adversary $B = (B_1, B_2)$ can be defined formally as bellow:

$B_1^{O_1(\cdot)}(pk)$
$\quad (x_0, x_1, s) \leftarrow A_1^{O_1(\cdot)}(pk); M \leftarrow \{x_0, x_1\};$
*return* $(M, s)$

$B_2^{O_2(\cdot)}(M, s, y)$
$\quad d \leftarrow A_2^{O_2(\cdot)}(x_0, x_1, s, y); v \leftarrow d$
$\quad f : \{x_0, x_1\} \rightarrow \{0,1\}$
$\quad f(x) = \begin{cases} 0 & \text{if } x = x_0 \\ 1 & \text{if } x = x_1 \end{cases}$
*return* $(v, f)$

Since $A_1$ and $A_2$ are poly-time algorithms, so are $B_1$ and $B_2$.

We define $p(b)$ and $p(b')$ as bellow:

$p(b) = Pr[(pk, sk) \xleftarrow{R} K(k); (M, s) \leftarrow A_1^{O_1(\cdot)}(pk); x_1 \leftarrow M; y \leftarrow E_{pk}(x_1);$
$x_0 \leftarrow Sample(M, s); (v, f) \leftarrow A_2^{O_2(\cdot)}(M, s, y) : v = f(x_b)]$
$= Pr[(pk, sk) \xleftarrow{R} K(k); (M, s) \leftarrow A_1^{O_1(\cdot)}(pk); x_1 \leftarrow M; y \leftarrow E_{pk}(x_1);$
$x_0 \leftarrow Sample(M, s); (v, f) \leftarrow A_2^{O_2(\cdot)}(M, s, y) : v = f(x_1)]$

$$p'(b) = Pr[(pk, sk) \xleftarrow{R} K(k); (x_0, x_1, s) \leftarrow A_1^{O_1(\cdot)}(pk); y \leftarrow E_{pk}(x_b);$$
$$d \leftarrow A_2^{O_2(\cdot)}(x_0, x_1, s, y) : d = 1]$$

Now, since the function $f$ is a deterministic one, the probability that $v = f(x_1)$ equals to the probability that $d = 1$. So $p(b)$ and $p'(b)$ will be equal and according to the definition of the advantage of the adversary we will have:

$$Adv_{P,A}^{css-atk}(k) = p(1) - p(0) = p'(1) - p'(0) = Adv_{P,A}^{ind-atk}(k)$$

So, whenever $Adv_{\Pi,A}^{ind-atk}(k)$ is not a non-negligible function, neither is $Adv_{P,A}^{css-atk}(k)$.

Now we proof the other side of the theorem, namely $IND\_ATK \Rightarrow CSS\_ATK$:

Our proof is again by contradiction.

Suppose $B = (B_1, B_2)$ is an adversary attacking semantic goal of the cryptosystem, i.e. can extract some non-negligible information about plaintext from its corresponding ciphertext. We construct the adversary $A = (A_1, A_2)$ to attack the indistinguishability goal of the cryptosystem by using $B = (B_1, B_2)$ as a subroutine.

The adversary $A = (A_1, A_2)$ is defined as bellow:

$A_1^{O_1(\cdot)}(pk)$
  $(M, s) \leftarrow B_1^{O_1(\cdot)}(pk); x_0, x_1 \leftarrow M;$
return $(x_0, x_1, s)$

$A_2^{O_2(\cdot)}(x_0, x_1, s, y)$
  $(v, f) \leftarrow B_2^{O_2(\cdot)}(M, s, y);$
  if $v = f(x_0)$ then $d \leftarrow 0;$
  if $v = f(x_1)$ then $d \leftarrow 1;$
  else $d \xleftarrow{R} \{0,1\};$
return $d$

Now if the function $f$ is a polynomially verifiable function, the algorithms $A_1$ and $A_2$ run in polynomial time.

Again for $b \in \{0,1\}$ we define $p(b)$ and $p(b')$ as bellow:

$$p(b) = Pr[(pk, sk) \xleftarrow{R} K(k); (x_0, x_1, s) \leftarrow A_1^{O_1(\cdot)}(pk); y \leftarrow E_{pk}(x_b);$$
$$d \leftarrow A_2^{O_2(\cdot)}(x_0, x_1, s, y) : d = 1]$$
$$p'(b) = Pr[(pk, sk) \xleftarrow{R} K(k); (M, s) \leftarrow A_1^{O_1(\cdot)}(pk); x_1 \leftarrow M; y \leftarrow E_{pk}(x_1);$$
$$x_0 \leftarrow Sample(M, s); (v, f) \leftarrow A_2^{O_2(\cdot)}(M, s, y) : v = f(x_b)]$$
$$= Pr[(pk, sk) \xleftarrow{R} K(k); (M, s) \leftarrow A_1^{O_1(\cdot)}(pk); x_1 \leftarrow M; y \leftarrow E_{pk}(x_1);$$
$$x_0 \leftarrow Sample(M, s); (v, f) \leftarrow A_2^{O_2(\cdot)}(M, s, y) : v = f(x_1)]$$

The algorithm $A_1$ outputs the value $1$ in the cases bellow:

If $v = f(x_1)$ and $v \neq f(x_0)$ the output will deterministically be $1$.

If $v \neq f(x_1)$ and $v \neq f(x_0)$ the output will be $1$ by the probability $1/2$.

Thus we have

$$p(1) = p'(1)(1-p'(0)) + \frac{1}{2}(p'(1)p'(0) + (1-p'(1))(1-p'(0))) = \frac{1}{2} + \frac{1}{2}(p'(1) - p'(0))$$

So,

$$Adv_{P,A}^{ind-atk}(k) = p(1) - p(0) = 2p(1) - 1 = p'(1) - p'(0) = Adv_{P,A}^{css-atk}(k)$$

Now whenever $Adv_{P,A}^{css-atk}(k)$ is not non-negligible function, so is $Adv_{P,A}^{ind-atk}(k)$.

## 5 Conclusion

In this paper, we formulated semantic security in comparison based framework. This unifies the definition frameworks of semantic security and indistinguishability and facilitates the study of relations between indistinguishability and semantic security against chosen plaintext attack, non-adaptive chosen ciphertext attack and adaptive chosen ciphertext attack removing the contradictions that was mentioned in the introduction. Then we suggested a simple proof for the equivalence of semantic security and indistinguishability in the new setting. Our proof is more understandable than previous ones and does not need any intermediate goals such as non-devidability to be defined for definition framework unification.